\documentclass[useAMS,usenatbib]{mn2e} \newif\ifdraft \draftfalse
\usepackage{graphics}
\usepackage{rotating}
\usepackage{amsmath}
\usepackage{amssymb}
\usepackage{color}

\newcommand{\vex}{\vspace{1ex}}

\renewcommand{\aa}{A\&A}
\newcommand{\aj}{AJ}
\newcommand{\apj}{ApJ}
\newcommand{\apjs}{ApJS}
\newcommand{\apss}{Ap\&SS}

\newcommand{\pasa}{PASA}
\newcommand{\mnras}{MNRAS}
\newcommand{\hm}{\hphantom{-}}

\newcommand{\atca}{\mbox{\sc atca-\small{104}}}
\newcommand{\ceduna}{\mbox{\sc ceduna}}
\newcommand{\tid}{\mbox{\sc dss\small{45}}}
\newcommand{\hobart}{\mbox{\sc hobart\small{26}}}
\newcommand{\mopra}{\mbox{\sc mopra}}
\newcommand{\parkes}{{\sc parkes}}
\newcommand{\lavlba}{{\sc la--vlba}}
\newcommand{\ovvlba}{{\sc ov--vlba}}
\newcommand{\SNR}{{\mbox{\rm SNR}}}

\newcommand{\getlength}[1]{\ifx#1\end \let\next=\relax
            \else\advance\count255 by1 \let\next=\getlength\fi \next}
\newcommand{\ifnularg}[1]{ \count255=0 \getlength#1\end \ifnum\count255=0 }
\newcommand{\ifm}{\makebox{}\ifmmode}
\long\def\ifundefined#1#2#3{\expandafter\ifx\csname
  #1\endcsname\relax#2\else#3\fi}
\newcommand{\beq}   { \begin{eqnarray} }
\newcommand{\eeq}[1]{ \ifnularg{#1} end{eanarray} \else
                      \label{#1}\end{eqnarray}    \fi }
\newcommand{\eeql}   { \end{eqnarray} }
\newcommand{\eeqn}   { \nonumber \end{eqnarray} }
\newcommand{\Frac}[2]{\frac{\displaystyle\strut #1}{\displaystyle\strut #2} }
\newcommand{\lp}{ \left(  }
\newcommand{\rp}{ \right) }
\newcommand{\dss}{\displaystyle}
\newcommand{\Cov}{ \mathop{ \rm Cov }\nolimits }
\newcommand{\un}[1]{\underline{#1}}
\newcommand{\nc}[1 ]{ \multicolumn{1}{c}{#1} }

\newcommand{\ntab}[2]{ \multicolumn{1}{#1}{#2} }

\newcommand{\nnntab}[2]{ \multicolumn{3}{#1}{#2} }

\newcommand{\Number}[1]{\ifnum#1<10\relax0\number#1\else\number#1\fi}
\newcommand{\isodate}{
\count151=\time
\divide\count151 by 60
\count151=\count151
\multiply\count151 by 60
\count152=\time
\advance\count152 by -\count151
\divide\count151 by 60
\count152=\count151
\multiply\count151 by 60
\count153=\time
\advance\count153 by -\count151
\Number{\year}.\Number{\month}.\Number{\day}--\Number{\count152}:\Number{\count153}
}
\definecolor{Dred}{rgb}{0.312,0.070,0.070}
\definecolor{Dblue}{rgb}{0.070,0.070,0.312}
\definecolor{Dgreen}{rgb}{0.070,0.312,0.070}
\definecolor{Db}{rgb}    {0.050,0.0,0.320}

\newcommand{\Blb}[1]{\textcolor{Dblue}{\bf #1}}

\newcounter{note}
\let\oldmarginpar\marginpar
\renewcommand\marginpar[1]{\-\oldmarginpar[\raggedleft\footnotesize #1]%
{\raggedright\footnotesize #1}}
\newcommand{\Note}[1]{\Rdb{#1}{\addtocounter{note}{1}%
\marginpar{\small\underline{\Rdb{Corr \arabic{note}}}}}}
\newcommand{\note}[1]{\Rdb{#1}}
\renewcommand{\note}[1]{#1}
\renewcommand{\Note}[1]{#1}
\volume{414}
\pubyear{2011}
\pagerange{2528--2538}
\title[The LBA Calibrator Survey --- LCS1]{The LBA Calibrator Survey
of southern compact extragalactic radio sources --- LCS1}
\author[Petrov et al.]{
  \parbox[t]{\textwidth}{
     Leonid Petrov$^{1}$\thanks{E-mail:Leonid.Petrov@lpetrov.net},
     Chris Phillips$^{2}$,
     Alessandra Bertarini$^{3}$,
     Tara Murphy$^{4,5}$, and
     \mbox{Elaine M.~Sadler}$^{4}$
  }
\vspace{1.0ex} \\
$^{1}$ADNET Systems, Inc./NASA GSFC, Code 610.2, Greenbelt, MD 20771 USA \\
$^{2}$CSIRO Astronomy and Space Science, PO Box 76, Epping, NSW 1710,
      Australia \\
$^{3}$Institute of Geodesy and Geoinformation, University of Bonn, Nussallee 17, Bonn, Germany and\\
      Max Planck Institute for Radioastronomy, Bonn, Germany \\
$^{4}$Sydney Institute for Astronomy, School of Physics, The University of Sydney, NSW 2006, Australia \\
$^{5}$School of Information Technologies, The University of Sydney, NSW 2006, Australia
}

\pagerange{\pageref{firstpage}--\pageref{lastpage}}

\begin{document}

\maketitle
\label{firstpage}

\begin{abstract}

\ifdraft
  \vspace{-58ex} \hfill \framebox{\Blb{\LARGE\bf Draft of \isodate}} \vspace{54ex}
\fi

   We present a catalogue of \note{accurate} positions and correlated flux
densities \note{for}
410 flat-spectrum, compact extragalactic radio sources previously
detected in the AT20G survey. The catalogue spans the declination range
$[-90\degr, -40\degr]$ and was constructed from four 24-hour VLBI observing
sessions with the Australian Long Baseline Array at 8.3~GHz. The \note{VLBI}
detection rate in these experiments is 97\%, the median uncertainty of
\note{the} source positions is 2.6~mas, \note{and} the median correlated
flux density \note{on projected baselines longer than 1000\,km} is 0.14~Jy.
The \note{goals of this work are} 1)~to provide a pool of \note{southern}
sources with positions \note{accurate to a few milliarcsec, which can be used}
for phase referencing observations, geodetic VLBI and space navigation;
2)~to extend the complete flux-limited sample of compact extragalactic sources
to the southern hemisphere; and 3)~to investigate \note{the} parsec-scale
properties of high-frequency selected sources from the AT20G survey.
As a result of \note{this VLBI campaign}, the number of compact
radio sources \note{south of declination $-40\degr$ which have
measured VLBI correlated flux densities and positions known to
milliarcsec accuracy has} increased by a factor of 3.5. The catalogue
and supporting material is available at {\tt http://astrogeo.org/lcs1}.

\end{abstract}


\section{Introduction}

   Catalogues of positions of compact extragalactic radio sources with
the highest accuracy are important for many applications. \Note{These include}
 imaging faint radio sources in the phase referencing mode, differential
astrometry, space geodesy, and space navigation. The method of VLBI
first proposed by \citet{r:mat65} allows us to derive the position of
sources with nanoradian precision (1 nrad $\approx$ 0.2~mas).
The first catalogue of source coordinates determined with VLBI
contained 35~objects \citep{r:first-cat}. Since then hundreds of sources
have been observed under geodesy and astrometry VLBI observing programs
at 8.6 and 2.3~GHz (X and S bands) using the Mark3 recording system at
the International VLBI Service for Geodesy and Astrometry (IVS) network.
Analysis of these observations resulted in the ICRF catalogue of
608~sources \citep{r:icrf98}.

\Note {The Very Long Baseline Array}
(VLBA) \Note {was later used to measure the positions of $~\!4000$ compact
radio sources} in the VLBA Calibrator Survey (VCS)
\citep{r:vcs1,r:vcs2,r:vcs3,r:vcs4,r:vcs5,r:vcs6}
and the geodetic program RDV \citep{r:rdv}. All sources with declinations
above $-45\degr$ detected using Mark3/Mark4 under IVS programs were
re-observed with the VLBA \Note{in the} VCS and RDV programs, which significantly
improved the accuracy of their positions. As a result of these efforts, the
probability of finding a calibrator with a VLBI-determined position
greatly increased. In the declination range $\delta > -30\degr$ the
probability of finding a calibrator within a $3\degr$ radius of
a given position reached 97\% by 2008.

\Note{Since} the VLBA is located in the northern hemisphere, observations
in the declination zone $[-50\degr, -30\degr]$ are difficult and the array
\Note{cannot observe} sources with $\delta < -52\degr$. In 2008, the probability
of finding a calibrator within a radius of $3^\circ$ was 75\% in the
declination zone $[-40^\circ, -30^\circ]$ and 42\% for declinations
\Note{south of }$-40^\circ$. The VLBI calibrator list\footnote{Available at
{\tt http://astrogeo.org/rfc}} in 2008 had 524~sources in the zone
$[+52^\circ, +90^\circ]$, \Note{but} only 98~objects in the zone
$[-52^\circ, -90^\circ]$ \Note{which cannot be reached by} the VLBA.
These southern sources
were observed during geodetic experiments and during two dedicated southern
hemisphere astrometry campaigns \citep{r:fey_south1,r:fey_south2}.
The reason for this disparity is the scarcity of VLBI antennas in the
southern hemisphere, particularly stations with dual frequency S/X
receivers and geodetic recording systems. Also until recently there
has been a lack of good all-sky \Note{catalogues suitable for finding
candidate VLBI calibrators}.

The Australian Long Baseline Array (LBA) consists of six
antennas located in Australia with the South Africa station {\sc hartrao}
often joining in. \Note{This VLBI network operates 3--4 observing sessions
a year, each about one week long}. Although the hardware used by the LBA
was not designed for geodesy and absolute astrometry observations, it was
demonstrated by \citet{r:geo_lba} that despite significant technical
challenges, absolute astrometry VLBI observations with the LBA network
is feasible. In a pilot experiment in June 2007, \Note{the} positions of
participating stations were determined with accuracies 3--30~mm, and
positions of \Note{five} new sources were determined with accuracies 2--5~mas.

  Inspired by these astrometric results, we launched the X-band LBA Calibrator
Survey observing campaign (LCS), with the aim of determining milliarcsecond positions
and correlated flux densities \Note{for} one thousand compact
extragalactic radio sources \Note{at declinations south of $-30^\circ$}. The overall
objective of this campaign is to match the density of calibrators in the
northern hemisphere and \Note{so} eliminate the disparity.

  We \Note{have three long-term goals in} this campaign. \Note{Firstly, setting up}
a dense grid of calibrators with precisely known positions within several degrees
of any target \Note{will make} make phase-referencing observations of weak targets \Note{feasible}.
According to \citet{r:wrobel_pr}, \Note{63\% of VLBA observations in 2003--2008}
were made in the phase referencing mode. A dense grid of calibrators also
makes {\note it possible to do} differential astrometry of Galactic objects such as pulsars
and masers, and allows direct determination of the parallax at distances
up to several kiloparsec \citep{r:del09}. These sources form the pools
of targets for observations under \Note{the geodesy programs} and for space
navigation.

  \Note{Our} second goal is to extend the complete flux-limited sample of
compact extragalactic radio sources (with emission from milliarcsecond-size
regions) to the entire sky. According to Kovalev (private communication, 2010),
analysis of the $\log N$--$\log S$ diagram of the VLBI calibrator list
suggests the sample  of radio--loud Active \Note{Galactic} Nuclei (AGN)
is complete at the level of correlated flux density 200~mJy at X-band
at spatial frequencies 25~$M\lambda$ at $\delta > -30\degr$.
Extending \Note{this} complete sample to the entire sky will make it
possible to generalize the properties of the sample, such as distribution
of compactness, distribution of brightness temperature, bulk motion, viewing
angle, irregularities of the spatial distribution, to the
entire population of radio loud AGN.

  The third goal \Note{is} to investigate \Note{the} properties of high-frequency
selected radio sources from \Note{the} Australia Telescope 20~GHz (AT20G) survey
\citep{r:at20g,r:mass10}. Obtaining observations of a subsample of AT20G
sources with milliarcsecond resolution will allow us to investigate \Note{the}
properties, such as spectral index, polarization fraction and variability,
of a population of extremely compact sources.

  In this paper, we present the results from the first four 24-hour experiments
observed in 2008--2009.  \Note{The selection of candidate sources from} the
AT20G catalogue is discussed in section~\ref{s:selection}. The station setups
during the observing sessions is described in section~\ref{s:observations}.
\Note{The} correlation and post-correlation analysis, which is rather different from
ordinary VLBI experiments, is discussed in sections~\ref{s:correlation} and
\ref{s:analysis}. \Note{An error analysis} of single-band observations, including
evaluation of ionosphere-driven systematic errors, is given in
subsection~\ref{s:errors}. The catalogue of source positions and correlated
flux densities is presented in section~\ref{s:catalog}, \Note{and} the results are
summarized in section~\ref{s:summary}.

\section{Candidate source selection}
\label{s:selection}

  The \Note{Australia Telescope 20\,GHz (AT20G) survey} is a blind radio survey
\Note{carried out} at 20 GHz with the Australia
Telescope Compact Array (ATCA) between 2004 and 2008 \citep{r:at20g}.
It covers the whole sky south of declination 0\degr. The source catalogue
is an order of magnitude larger than previous catalogues of high-frequency
radio sources, with 5890 sources above a 20 GHz flux-density limit of
40~mJy.
All AT20G
sources have total intensity and polarization measured at 20~GHz, and
most sources south of declination $-15\degr$ also have near-simultaneous
flux-density measurements at 4.8 and 8.6~GHz. A total of 1559 sources were
detected in polarised total intensity at one or more of the three frequencies.
There are also optical identifications and redshifts for a significant
fraction of the catalogue.

  This high-frequency catalogue provides a good starting point for selecting bright,
compact sources and candidate calibrators. \citet{r:mass10} show that almost
all the flat-spectrum AT20G sources with spectral index $\alpha>-0.5$
are unresolved on scales of 0.1--0.2\,arcsec in size at 20\,GHz
(see their fig.~3.4). The few exceptions are either (i) foreground
Galactic and LMC sources such as planetary nebulae, HII regions and pulsar
wind nebulae, which have a flat radio spectrum due to their thermal emission but
are usually resolved on scales of a few arcsec \citep{r:at20g}, or (ii)
flat-spectrum extragalactic sources which are gravitationally-lensed,
like PKS\,1830$-$211.

  For 64\% of sources from the AT20G catalogue, flux densities were determined
in three bands, 5.0, 8.4 and 20~GHz. We used these measurements
to calculate the spectral index $\alpha$ \Note{($F \propto f^\alpha$)}.
We selected a set of 684 objects, \Note{not previously} observed with VLBI,
\Note{which had a flux density $> 150$~mJy at 8.3~GHz and a} spectral index $> -0.6$.
We then split this sample into two subsets: 410 high priority sources with
flux densities $> 200$~mJy and spectral indices $> -0.5$, and all others.
In addition, we selected 14 flat-spectrum objects that did not have flux
density measurements at 5~GHz and 8~GHz in AT20G. Their spectral indices
were determined by analyzing historical single dish observations found in
the Astrophysical CATalogs support System CATS \citep{r:cats} \Note{database,
which by 2010 included data from 395 catalogues from radioastronomy surveys.}

  In addition to \Note{these} target sources, we identified a set of 195 sources that
we used for calibrator selection. These were bright sources previously
observed at the VLBA or IVS network with position known with accuracies
better 0.5~mas and  with correlated flux densities $> 500$~mJy
on baselines longer than 5000 km.

\subsection{The Long Baseline Array}

\begin{figure}
   \includegraphics[width=0.48\textwidth]{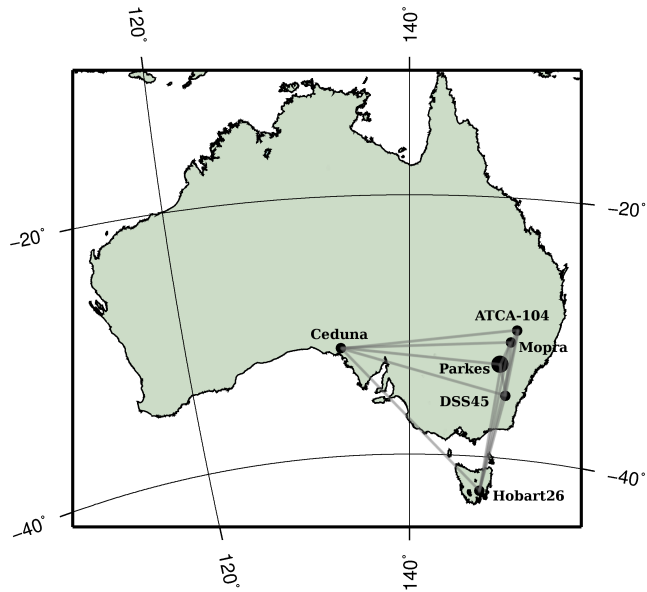}
   \caption{The LBA network used for LCS1 observations.}
   \label{f:lba_map}
\end{figure}

   The LBA network used for LCS1 consists of 6 stations:
\atca, \ceduna, \tid, \hobart, \mopra\ and \parkes\
(Fig.~\ref{f:lba_map}, Table~\ref{t:lba}). The maximum baseline length
of the network is 1703~km, the maximum equatorial baseline  projection
is 1501~km, and the maximum polar baseline projection is 963~km.

\begin{table}
    \caption{The LBA network stations. The average System Equivalent Flux
             Density (SEFD) at 8.3~GHz at elevation angles $> 60\degr$
             achieved in LCS1 experiments \note{is shown} in the last column.}
    \begin{tabular}{l @{\qquad} l @{\qquad}  l @{\qquad}  l @{\qquad}  r}
       \hline
       Name            &  \ntab{c}{$\phi_{gc}$}   &
                          \ntab{c}{$\lambda$}     & Diam & SEFD \\
       \hline
       \atca     & -$30\degr.15$  & $149\degr.57$ & 22 m & 350 Jy  \\
       \ceduna   & -$31\degr.70$  & $133\degr.81$ & 32 m & 490 Jy  \\
       \tid      & -$35\degr.22$  & $148\degr.98$ & 34 m & 120 Jy  \\
       \hobart   & -$42\degr.62$  & $147\degr.44$ & 26 m & 620 Jy  \\
       \mopra    & -$31\degr.10$  & $149\degr.10$ & 22 m & 380 Jy  \\
       \parkes   & -$32\degr.82$  & $148\degr.26$ & 64 m &  36 Jy  \\
       \hline
    \end{tabular}
    \label{t:lba}
\end{table}

\subsection{Observation scheduling}

  Observation schedules were prepared with the software program {\sf sur\_sked}.
The scheduling goal was to observe each target source \Note{with} all antennas
of the array in \Note{three} scans of 120~seconds each in the first three experiments
and in \Note{four} scans in the last experiment. After tracking a target source for 120
seconds, each antenna immediately slewed to the next object. The minimum time
interval between consecutive scans of the same source was 3~hours.
The scheduling software computed \Note{the} elevation and azimuth of
\Note{each} candidate
source, with constraints on the horizon mask and maximum elevations to which antennas
can point. It calculated the slewing time taking into account the antenna
slewing rate, slewing acceleration and cable wrap constraints. The sources
were scheduled \Note{to minimize} slewing time and \Note{fit the}
scheduling constraints. A score was assigned for each target source visible
at a given time, \Note{depending} on the slewing time, the history of past
observations in the experiment and the amount of time that the source would
remain visible to all antennas in the array.

   Every hour a set of \Note{four} calibrator sources was observed: two scans
\Note{where all antennas had an} elevation in the range  $[12\degr, 45\degr]$,
one scan at elevations $[32\degr, 45\degr]$ and one scan at elevations
$[45\degr, 90\degr]$. Scans with calibrator sources were 70 seconds long.
The scheduling algorithm for each set found all combinations
of calibrator sources that fell in the \Note{right} elevation range, and selected
the sequence of four objects that minimized the slewing time. The purpose
of these observations was 1)~to serve as amplitude and bandpass calibrators;
2)~to improve \Note{the} robustness of estimates of the path delay in the neutral
atmosphere; and 3)~to tie \Note{the} positions of new sources to existing
catalogues such as the ICRF catalogue \citep{r:icrf98}.

  The scheduling algorithm assigned \Note{four} scans for 18\% of target sources,
\Note{three} scans for 76\% \Note{of} sources, \Note{two scans for 3\% of} sources,
and \Note{one scan for 3\% of} sources. \Note{Of the 26 targets with fewer than
three scans} in a given observing session, 18~were observed in two experiments.
During a 24~hour experiment, 11.5--12.0 hours were used for observing target
sources, 1.5--2.0 hours for observing \Note{calibrators}.

\section{Observations}
\label{s:observations}

\begin{table}
    \caption{Observation Summary for each observing session}

    \par\hspace{-1.8em}
    \begin{tabular}{@{\hspace{0.0em}}cllc}
       \hline
       Telescope            & Recorder  & Polarization
       &  \parbox{0.16\textwidth}{\centering Frequency  Bands (MHz)} \\
       \hline
       \nnntab{l}{\hspace{-0.4em}\un{\sf Experiment v254b, February 05 2008}} \vex\vex  \\
       \parbox{0.06\textwidth}{
                                \parkes \\
                                \atca   \\
                                \mopra  \\
                                \hobart \\
                                \ceduna
                              } & LBADR  &
         RCP only   & \parbox{0.16\textwidth}{8256--8272  8272--8288 \\
                                              8512--8528 8528--8544   }
                      \vex \vex \vex \vex \\

       \nnntab{l}{\hspace{-0.4em}\un{\sf Experiment v271a, August 10 2008}} \vex  \\
       \parbox{0.06\textwidth}{
                                \parkes \\
                                \hobart \\
                                \tid
                              } & Mark5  &
       RCP only   & \parbox{0.16\textwidth}
                    {8200--8216  8216--8232 \\
                     8232--8248  8248--8264 \\
                     8456--8472  8472--8488 \\
                     8488--8504  8504--8520
                    } \vex\vex\vex \\

       \ceduna & LBADR  &  RCP only & \parbox{0.16\textwidth}
                                       {8200--8216  8216--8232 \\
                                        8456--8472 8472--8488}
                                       \vex \vex \vex \\

       \parbox{0.06\textwidth}{
                                \atca \\
                                \mopra
                              }  & LBADR  &  RCP \& LCP &
       \parbox{0.16\textwidth}{8200--8216  8216--8232 \\
                               8456--8472 8472--8488}
            \vex \vex \vex \vex \\

       \nnntab{l}{\hspace{-0.4em}\un{\sf Experiment v271b, November 28 2008}} \vex  \\
       \parbox{0.06\textwidth}{
                                \parkes \\
                                \hobart \\
                                \tid
                              } & Mark5  & RCP only   &
                              \parbox{0.16\textwidth}{8200--8216  8216--8232  \\
                                                      8264--8280  8328--8344  \\
                                                      8392--8408  8456--8472  \\
                                                      8472--8488  8520--8536 }
                              \vex \vex \vex \\

        \ceduna & LBADR  &  RCP only & \parbox{0.16\textwidth}
                                                 {8200--8216  8216--8232  \\
                                                  8456--8472  8472--8488}
                              \vex \vex \vex \\
        \parbox{0.06\textwidth}{
                                 \atca \\
                                 \mopra
                               }  & LBADR  &  RCP \& LCP &
                               \parbox{0.16\textwidth}{8200--8216  8216--8232    \\
                                                       8456--8472  8472--8488}
                               \vex \vex \vex \vex \\

       \nnntab{l}{\hspace{-0.4em}\un{\sf Experiment v271c, July 04 2009}} \vex  \\

       \parbox{0.06\textwidth}{
                                \hobart
                              } & Mark5  &
         RCP only   & \parbox{0.16\textwidth}{8200--8216  8216--8232  \\
                                              8232--8248  8328--8344  \\
                                              8344--8360  8456--8472  \\
                                              8472--8488 8488--8504 }
                      \vex \vex \vex \\
       \parbox{0.06\textwidth}{
                                \parkes \\
                              } &
                              \parbox{0.08\textwidth}{ Mark5 \& \\
                                                       LBADR
                                                     } &
         RCP \& LCP & \parbox{0.16\textwidth}{8200--8216  8216--8232  \\
                                              8232--8248  8328--8344  \\
                                              8344--8360  8456--8472  \\
                                              8472--8488 8488--8504 }
                      \vex \vex \vex \\
         \ceduna    & LBADR  &  RCP only &
                            \parbox{0.16\textwidth}{8200--8216  8216--8232 \\
                                                    8456--8472 8472--8488}
                      \vex \vex \vex \\
        \parbox{0.06\textwidth}{
                                 \atca \\
                                 \mopra
                               }  & LBADR  &  RCP \& LCP &
                               \parbox{0.16\textwidth}{8200--8216  8216--8232  \\
                                                       8456--8472  8472--8488}
                              \\
       \hline
    \end{tabular}
    \label{t:obssum}
\end{table}

  As mentioned in the introduction, not all of the telescopes in the LBA
network are capable of observing in the typical geodetic/astrometric
mode of dual S/X frequency bands with multiple spaced subbands and
recorded to a Mark5 VLBI system.

  \Note{The} \hobart, \parkes, and \tid\ have a Mark5 recording system and
a Mark4 \citep{r:mark-4} baseband conversion rack. \Note{However for} the
first epoch the LBADR system  \citep{r:lbadr} was used. \Note{For} all
subsequent experiments the Mark5 system was used. Data were recorded onto
normal Mark5 diskpacks and shipped to the Max-Plank Institut f\"{u}r
Radioastronomie in Bonn for processing.

The \atca, \mopra\ and \ceduna\ only have the standard LBA VLBI backend
consisting of an Australia Telescope National Facility (ATNF) Data
Acquisition System (DAS) with an LBADR recorder. The
ATNF DAS only allows two simultaneous intermediate frequencies (IFs):
either 2 frequencies or 2 polarizations. For each of these IFs the input
64~MHz analog IF is digitally filtered to 2 contiguous 16~MHz bands.
\atca, \mopra, and \parkes\ have two ATNF DAS, however the \Note{IF conversion
system at each telescope means it is impractical to run in any
modes other than 2 frequencies} and dual polarization. The LBADR data
format is not compatible with the Mark4 data processor at Bonn.
A custom program was written to translate the data to Mark5B format then
it was electronically copied using the Tsunami-UDP application
to Bonn, before being copied onto Mark5 diskpacks.

  In the last experiment \parkes\ recorded on the LBADR system in parallel
with the Mark5. As this included left circular polarization (LCP) which was
not \Note{recorded} on the Mark5 system and since \atca\ and \mopra\ recorded
both RCP and LCP, the LBADR data were also \Note{sent} to Bonn and LCP
correlated against \atca\ and \mopra.

  The main limiting factor for frequency selection was the LBADR backend
at \atca, \mopra\ and \ceduna\ where a setup with bands centered
on $\sim$8.2~and$\sim$8.5~GHz was chosen. For the telescopes with Mark5
recorders, the setup was chosen to overlap in frequency with the LBADR setup
but also including more frequencies to gain sensitivity. The frequency setup
was changed between experiments in order to explore the \Note{feasibility
of recording} at 512~Mbit/s at those stations that can support it.
The setup for each observing experiment is described in Table~\ref{t:obssum}.

  The Australia Telescope Compact Array consists of six 22~metre antennas
and may observe as a single antenna or as a phased array of 5
dishes\footnote{The sixth dish is located at the fixed pad far away from
other antennas, which makes its phasing with the rest of the compact
array too difficult.}. Since no tests of using a phased ATCA array as an
element of the VLBI network for absolute astrometry observations were made
before 2010, we decided to use a single dish of the ATCA in order to
avoid \Note{the} risk of introducing unknown systematic errors.

   The NASA Deep Space Network station \tid\ observed only 4.5 hours in v271a
and 6.5 hours in v271b during intermissions between receiving the telemetry
from Mars orbiting spacecraft.

\section{Correlation}
\label{s:correlation}

  The Bonn Mark4 Correlator was chosen for the correlation of these
experiments for two primary reasons. Firstly, the correlator was extensively
tested for use in space geodesy and absolute astrometry mode during
2000--2010 and is known to produce highly reliable group
delays. Secondly, it was equipped with four Mark5B units and eight
Mark5A units, which was a convenient combination since data
from \parkes\ were recorded in the Mark5B format and data from
\atca, \mopra, and \ceduna originally recorded in LBADR format were
transformed to the Mark5B format before correlation.

  Preparation for correlation required about two months due to
complications that arose from the scheduling of different patching and
channel outputs (fan out) for use at the stations.  The stations that had
Mark4 data acquisition racks delivered detailed log files, from
which we could reconstruct the channels used, but the LBA stations did not
provide log files and the track assignments had to be searched for by trial
and error, which was a time-consuming process. The absence of log files
also required some custom programming to reconstruct the log VEX file (lvex),
which the Mark IV correlator required to perform the correlation.
Further, the setup that was chosen was incompatible with the capabilities of
the hardware correlator and required an extensive work-around at the correlator.

  We chose to correlate with a window of 128~lags (corresponding to a delay
window width of \Note{8~$\mu$s) instead of the 32~lags (2~$\mu$s)} normally
used for stream correlation. \Note{This was} to allow for potentially large clock offsets due
to instrumental errors, and for large a~priori source coordinate errors.
The integration time was chosen to be as small as possible (0.5~s to 1.0~s)
to allow for potentially large residual fringe rates due to source
position errors.

  The Mark4 correlator can cope with only four stations simultaneously
when correlating 128~lags with short integration time. Since
the experiment was observed with five to seven stations in right circular
polarization (RCP) and two stations (\atca\ and \mopra)  also in
left circular polarization (LCP), the correlation had to be split in
passes.

  As a by-product of the multi-pass correlation, some baselines were
correlated more than once and this redundancy was exploited to select
the correlation with the best \SNR\ for each duplicated scan. For the LCP
correlation, only one pass was required since only the LBA stations
recorded LCP. {\sc parkes} recorded LCP only during v271c and this
was enabled by the use of two backends (Mark4 and LBA) in parallel.
Those stations were re-correlated in a second pass to produce all four
polarization products (left-left, right-right, left-right and
right-left). For this pass we had to restrict the number of lags to 32
due to correlator hardware constraints, but the integration time was
kept at 0.5~s.

  Fringe fitting was performed at the correlator using software
program {\sc fourfit}, the baseline-based fringe fit offered within
the Haystack Observatory Package Software ({\sc hops}) to estimate
the residual delay upon which the post-correlation analysis was based.
Inspection of fringe-fitted data is a convenient means of data quality
control, as one can examine the correlated data on a scan-by-scan basis,
looking for problems that might have occurred during correlation or
at the stations and permits flagging of bad data immediately.

\section{Data analysis}
\label{s:analysis}

  The correlator \Note{either} generates the spectrum of the cross-correlation function
\Note{directly,} or it can be easily derived from its raw output.
The spectrum was processed with the software program {\sc fourfit}, \Note{which}
for each scan and each baseline determined \Note{the} phase delay rate, narrow-band
delay and wide-band group delay (sometimes also called multi-band delay)
that maximized the fringe amplitude. \citet{r:ksp} \Note{give} a detailed
description of the fringe searching process and the distinction between
narrow-band and wide-band group delays. The wide-band delay is more precise
than narrow-band delay. The formal uncertainties of these delays,
$\sigma_n$ and $\sigma_w$ are computed by {\sc fourfit} the following way:
\beq
     \begin{array}{lcl}
          \sigma_n & = & \Frac{\sqrt{12}}{2\pi \Delta f \: \SNR} \\
          \sigma_w & = & \Frac{1}{2\pi \sigma_f \: \SNR} ,
     \end{array}
\eeq{e:e1}
    where $\Delta f$ the IF bandwidth, $\sigma_f$ the dispersions of
cyclic frequencies across the band, and \SNR\ is the ratio  of the fringe
amplitude from the wide band fringe search to the rms of the thermal noise.
The ratio of $\sigma_w/\sigma_n$ was in the range 27--28 for the LCS1
experiments, \Note{which} gives for observations with typical \SNR=30
$\sigma_w \approx 40$~ps and $\sigma_n$ around one nanosecond.

   \Note{Of the} 421 observed targets, 410 were detected. The
list of 11 \Note{undetected sources is given} in
Table~\ref{t:nd_lcs1}. In addition 111 calibrators were observed, all of which
were detected.

\begin{table}
    \caption{The list of target sources that were not detected in LCS1
             VLBI observations. The last three data columns contain
             the flux density in mJy extrapolated to 8.3~GHz (F), the
             spectral index ($\alpha$), and comment (C). Value \mbox{-9.99}
             means the spectral index was not available. Comments:
             1)~Not in AT20G catalogue; 2)~Flagged as an LMC source in
             AT20G catalogue; 3)~Galactic planetary nebula; 4)~Flagged
             as an extended source in AT20G catalogue. \note{The table is also
             available in the electronic attachment.}}
   \label{t:nd_lcs1}
   \par\hspace{-2em}
   \begin{tabular}{@{\!}l@{\:\:\:}l  @{\quad}
                   l@{\:\:}l@{\:\:}l @{\:\:\:}
                   l@{\:\:}l@{\:\:}l @{\quad}
                   r@{\:\:}r@{\:\:\:}r@{\hspace{-0.05em}}}
      \hline
      J2000-name & \ntab{c}{B1950} & \nnntab{c}{RA}     & \nnntab{c}{Dec} &
                   \ntab{c}{F}     & \ntab{c}{$\alpha$} & C \\
      \hline
      J0404$-$7109 & 0404$-$712 & 04 & 04 & 00.99 & -71 & 09 & 09.7  &  172 &  -0.57 & 1 \\
      J0538$-$6905 & 0539$-$691 & 05 & 38 & 45.66 & -69 & 05 & 03.1  &  204 &  -9.99 & 2 \\
      J0552$-$5349 & 0551$-$538 & 05 & 52 & 36.18 & -53 & 49 & 32.4  &  182 &  -0.34 & 1 \\
      J0938$-$6005 & 0937$-$598 & 09 & 38 & 47.20 & -60 & 05 & 28.7  &  196 &  -0.08 & 3 \\
      J0958$-$5757 & 0956$-$577 & 09 & 58 & 02.92 & -57 & 57 & 42.6  &  383 &   0.21 & 3 \\
      J1100$-$6514 & 1058$-$649 & 11 & 00 & 20.09 & -65 & 14 & 56.4  &  179 &  -0.08 & 3 \\
      J1150$-$5710 & 1147$-$569 & 11 & 50 & 17.87 & -57 & 10 & 56.0  &  349 &   0.23 & 3 \\
      J1325$-$4302 & 1322$-$427 & 13 & 25 & 07.35 & -43 & 02 & 01.8  &  327 &  -9.99 & 4 \\
      J1353$-$6630 & 1350$-$662 & 13 & 53 & 57.05 & -66 & 30 & 50.3  &  370 &   0.04 & 3 \\
      J1505$-$5559 & 1502$-$557 & 15 & 05 & 59.17 & -55 & 59 & 16.2  &  217 &  -0.06 & 3 \\
      J1656$-$4014 & 1653$-$401 & 16 & 56 & 47.53 & -40 & 14 & 24.4  &  427 &  -9.99 & 4 \\
      \hline
   \end{tabular}

\end{table}

\subsection{Data analysis: source position determination}

  The most challenging part of data analysis was resolving group delay
ambiguities. The algorithm for fringe fitting implemented in {\sc fourfit}
searches for a global maximum in the Fourier transform of the
cross-correlated function, averaged over individual IFs after correcting
phases for a fringe delay rate and a narrow-band group delay. Fringe
spectrum folding results in a rail of maxima of the Fourier transform of
the cross-correlated function with exactly the same amplitude and with
spacings reciprocal to the minimum difference in intermediate
\Note{frequencies, $1/16.0 \cdot 10^{6}\,\mbox{Hz} = 62.5$~ns. Within
one half of the 62.5~ns range, the fringe spectrum has several strong
secondary maxima. The second maximum at $1/256.0 \cdot 10^{8}\,\mbox{Hz}
\approx 3.9$~ns has the amplitude 0.981 of the main maximum, the third maximum
at $2/256.0 \cdot 10^{8}\,\mbox{Hz} \approx 7.8$~ns has the amplitude 0.924.
This high level of secondary maximum amplitudes is due to our
choice of intermediate frequencies that was determined by the hardware
limitations. Due to the presence of noise in the cross-correlated function,
the difference between the fringe amplitude at the main maximum and
at the secondary maxima is random, and therefore, the group delay is
determined with an uncertainty $\pm 1-3$ of the spacing between the
secondary maxima, 3.9~ns.}

  It should be noted that typical group delay ambiguity spacings
in geodetic observations are in the range 50--200~ns. For \Note{successfully}
resolving ambiguities, as a rule of thumb the predicted delay should be known with
an accuracy better than 1/6 of the ambiguity spacing, 600~ps in our case.
This is a challenge. The random errors of the narrow-band group delays are
too high \Note{to use} directly for resolving ambiguities in wide-band
group delays (see Fig.~\ref{f:sb_res}). In the framework of traditional
data analysis of geodetic VLBI observations, the maximum uncertainty in
prediction of the path delay is due to a lack of an adequate model for the
path delay in the neutral atmosphere. These errors are in the range of
300--2000~ps. The path delay through the ionosphere at 8.3~GHz is in the
range 30--1000~ps. In addition, an error $1''$  in a~priori source
position \Note{(a typical AT20G position error) causes an error in a~priori
time delay of up to 30~ns on a 1700\,km baseline.}

\subsubsection{Group delay ambiguity resolution}

  However, it is premature to conclude that resolving
group delay ambiguities is impossible. Firstly, we
need to use a state-of-the art a~priori model. Our computation
of theoretical time delays in general follows the approach of
\citet{r:masterfit} with some refinements. The most significant
ones are the following. The advanced expression for time delay
derived by \citet{r:Kop99} in the framework of general relativity
was used. The displacements caused by the Earth's tides were computed
using the numerical values of the generalized Love numbers presented
by \citet{r:mat01} following a rigorous algorithm described by
\citet{r:harpos} with a truncation at a level of 0.05~mm.
The displacements caused by ocean loading were computed by convolving
the Greens' functions with ocean tide models. The GOT99.2 model of diurnal
and semi-diurnal ocean tides~\citep{r:got99}, the NAO99 model \citep{r:nao99}
of ocean zonal tides, the equilibrium model \citep{r:harpos} of the pole tide,
and the tide with period of 18.6 years were used. Station displacements
caused by the atmospheric pressure loading were computed by convolving
the Greens' functions that describe the elastic properties of the
Earth \citep{r:farrell} with the output of the  atmosphere NCEP Reanalysis
numerical model \citep{r:ncep}. The algorithm of computations is described
in full detail in \citet{r:aplo}. The displacements due to loading caused
by variations of soil moisture and snow cover in accordance with GLDAS Noah
model \citep{r:gldas} with a resolution $0.25^\circ \times 0.25^\circ$
were computed using the same technique as the atmospheric pressure loading.
The empirical model of harmonic variations in the Earth orientation parameters
{\tt heo\_20101111} derived from VLBI observations according to the method
proposed by \citet{r:erm} was used. The time series of UT1 and polar motion
derived by the NASA Goddard Space Flight Center operational VLBI solutions
were used a~priori.

  The a~priori path delays in the neutral atmosphere in the direction of observed
sources were computed by numerical integration of differential equations of
wave propagation through the heterogeneous media. The four-dimensional field
of the refractivity index distribution was computed using the atmospheric
pressure, air temperature and specific humidity taken from the output of the
Modern Era Retrospective-Analysis for Research and Applications
(MERRA) \citep{r:merra}. That model presents the atmospheric parameters at
a grid $1/2\degr \times 2/3\degr \times 6^h$ at 72 pressure levels.

\begin{figure}
   \includegraphics[width=0.48\textwidth]{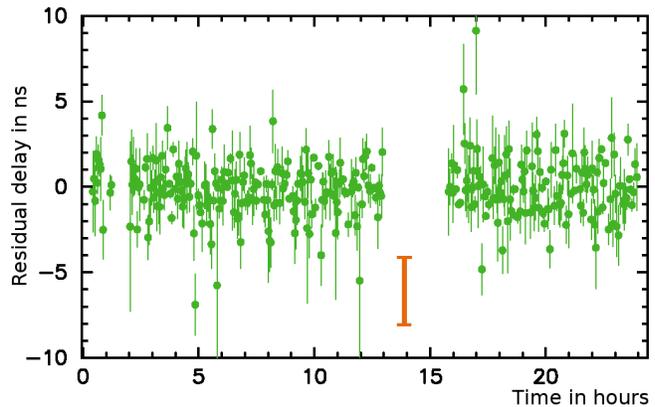}
   \caption{The post-fit residuals of narrow-band group delay
            LSQ solution at the baseline \ceduna/\hobart\ in experiment
            v271a on February 05, 2008. The bar in the middle of
            the plot corresponds to the ambiguity spacing of
            the wide-band path group delay.
           }
   \label{f:sb_res}
\end{figure}

  Secondly, we made a least square (LSQ) solution using narrow-band delay.
\Note{The positions of all target sources were estimated, as well as clock
functions for all stations except the one taken as a reference
and the residual atmosphere path delay in zenith directions. The clock function
was modeled as a sum of the 2nd degree polynomial over the experiment and
B-spline of the first order with the time span 60 minutes.  The residual
atmosphere path delay in zenith direction was modeled with B-spline of the
first order with time span 60~minutes. Constraints on rate of change
of clock function and atmosphere path delay were imposed}. After the removal
of outliers (1--2\% of points), the weighted root mean square (wrms)
of residuals was 1--4~ns. An example of narrow-band postfit residuals
is shown in Fig.~\ref{f:sb_res}.

\begin{figure}
   \includegraphics[width=0.48\textwidth]{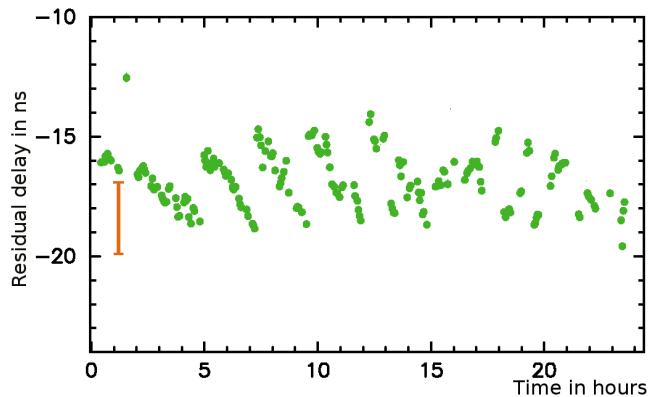}
   \caption{The wide-band a~priori group delays after substitution
            of adjustments to clock function, atmosphere path delay
            in zenith direction and source positions from the
            narrow-band LSQ solution at the baseline \ceduna/\parkes\
            in experiment v271a on February 05, 2008. The bar in
            the left side of the plot corresponds to the ambiguity
            spacing of the wide-band path group delay.
           }
   \label{f:gr_res}
\end{figure}

  The next step was to substitute the adjustments to clock function,
atmosphere path delay in zenith direction and source positions
from the narrow band LSQ solution to the differences between the wide-band
path delays and the theoretical group delays. An example of enhanced
wide-band group delays after the substitution and group delay ambiguity
resolution is shown in Fig.~\ref{f:gr_res}.

  It is still difficult to resolve ambiguities at long baselines,
but relatively easy to resolve the ambiguities at the inner part of the
array: \atca, \tid, \mopra, \parkes. The group delay ambiguity resolution
process starts from the baseline with the least scatter of a~priori
wide-band delays. After group delay ambiguity resolution at the
first baseline and temporary suppression \Note{of} observations with
questionable ambiguities, the LSQ solution with mixed delays is made:
wide-band group delays at baselines with resolved ambiguities and narrow-band
group delays at other baselines. In addition to other parameters,
baseline-dependent clock misclosures are estimated.

  In the absence of instrumental delays, the differences between wide-band
and narrow-band delays would be the zero mean Gaussian random noise.
Instrumental delays in the analogue electronics cause systematic changes
of these differences in time. Fortunately, these systematic changes are
smooth and small enough to allow solving for ambiguities with spacings as
small as 3.9~ns. Since the scatter of postfit residual of wide-band group
delays is a factor of 10--50 less than the ambiguity spacings, when the
instrumental delay is determined, the ambiguities can be resolved. Therefore,
the feasibility of resolving wide-band group delays hinges upon the
accuracy of determination of the differences between narrow-band and
wide-band group delays from analysis of narrow-band delays.

\begin{figure}
   \includegraphics[width=0.48\textwidth]{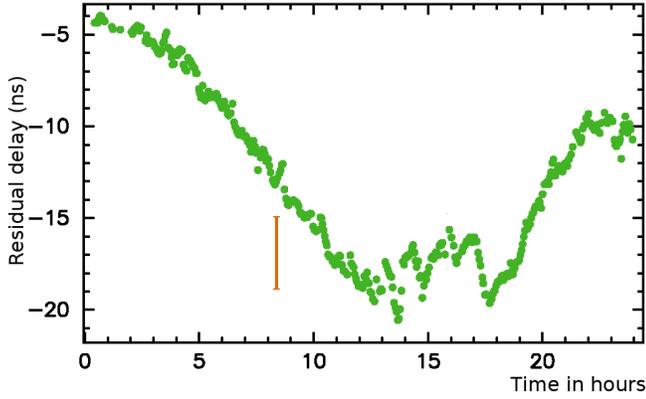}
   \caption{The wide-band a~priori group delays at the baseline
            \ceduna/\parkes\  in experiment v271a on February 05, 2008
            after substitution of adjustments to clock function,
            atmosphere path delay in zenith direction from the
            narrow-band LSQ solution and source positions from the
            wide-band LSQ solution at other baselines and group
            delay ambiguity resolution. The bar in the central
            side of the plot corresponds to the ambiguity
            spacing of the wide-band path group delay.
            }
   \label{f:sb_dif}
\end{figure}

  An example of the a~priori wide-band group delay with
the substituted adjustments to sources positions found from the wide-band
group delay solution at other baselines is shown in Fig.~\ref{f:sb_dif}.
After resolving ambiguities for the most of observations, the points
that were temporarily suppressed were restored and used in the solution.

\subsubsection{Ionosphere path delay contribution}

  Single band VLBI data are affected by a variable path delay through
the ionosphere. We attempted to model this path delay using GPS TEC
maps provided by the CODE analysis center for processing Global Navigation
Satellite System data \citep{r:scha98}. Details of computing the VLBI
ionospheric path delay using TEC maps from GPS
are given in \citet{r:vgaps}. The model usually recovers over 80\% of the
path delay at baselines longer \Note{than} several thousand kilometres.
However, applying
the ionosphere TEC model to processing LCS1 experiment did not improve the
solution and even degraded the fit. The same model and software
program certainly improved the fit and improved results when applied
to processing observations on intercontinental baselines.

  In order to investigate the applicability of the reduction for the path
delay in the ionosphere based on GPS TEC maps, we processed 29 IVS
dual-band geodetic experiments that included the 1089 km long baseline
\hobart/\parkes. We ran three solutions that included the data only
at this baseline. The first reference solution used ionosphere-free linear
combinations of X-band and S-band observables. The second solution used
X-band only group delays. The third solution used X-band only data and
applied the reduction for the path delay in the ionosphere from GPS TEC maps.

\begin{figure}
   \par\noindent
   \includegraphics[width=0.48\textwidth]{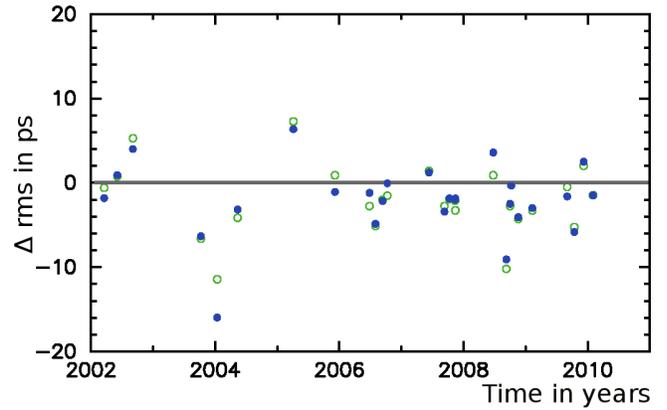}
   \caption{The difference in rms of postfit residuals
            with respect to the reference dual band X/S solution
            for two trial solutions: X-band only data (solid discs)
            and X-band data with ionosphere path delay from GPS TEC
            map applied (circles).
           }
   \label{f:delta_rms}
\end{figure}

  We discarded two experiments: one had a clock break at \hobart\
and another had the rms of postfit residual a factor of 5 greater than
usually due to a warm receiver. The rms differences of postfit residuals
with respect to the reference dual band X/S solution for two trial
solutions are shown in Fig.~\ref{f:delta_rms}: one with X-band only data
and another with X-band data with ionosphere path delays from GPS TEC map
applied. Considering the reference solution based on ionosphere-free group
delays as the ground truth, we expected that the rms differences between
the  X-band only solution (denoted with circles in Fig.~\ref{f:delta_rms})
be positive, and the rms differences of the X-band solution with the
ionosphere path delay from GPS TEC maps applied be also positive but
less than the rms differences between the X-band only solution. Instead, we
see that the rms of postfit residuals of the X-band only solution are
{\it less} than the rms of dual-band reference solution and applying
the ionosphere path delay from GPS TEC maps does not affect the rms
significantly.

\begin{figure}
   \includegraphics[width=0.48\textwidth]{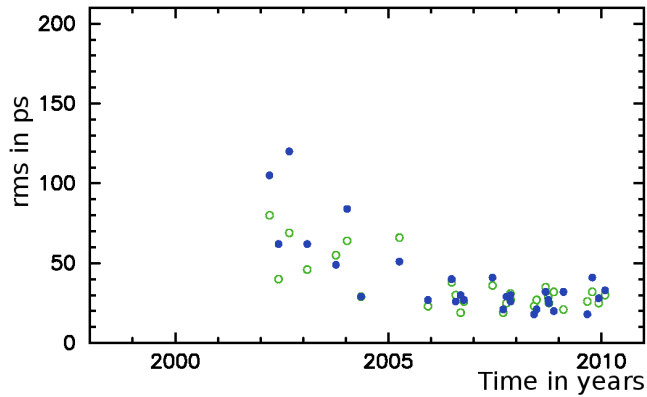}
   \caption{The rms of the ionosphere contribution from dual-band
            X/S observations (solid disks) and the rms of the differences
            of the path delay in the ionosphere from the dual-band
            X/S observations at the baseline \hobart/\parkes\
            and the GPS TEC ionosphere maps (circles).
           }
   \label{f:rms_iono_ho_pa}
\end{figure}

  We then computed the rms of the contribution of the ionosphere path
delay from dual-band X/S observations and the rms of the differences
between the path delay in the ionosphere from the dual-band
X/S observations and the GPS TEC ionosphere maps. They are shown
in Fig.~\ref{f:rms_iono_ho_pa}. We see that in 2002--2004 the ionosphere
path delay from GPS TEC maps was coherent with the dual-band ionosphere
path delay estimate. In 2005--2010 the ionosphere path delay was small with
rms $\sim\!\! 30$~ps and the ionosphere path delay from the GPS TEC model
was only partially coherent with the ionosphere path delay estimates from
dual-band VLBI observations. The baseline length repeatability, defined
as the rms of baseline length estimates after \Note{subtracting} the secular
drift due to tectonics, is the minimum when ionosphere-free linear combinations
of X/S observables are used: 17.0~mm, grows to 17.4~mm when X-band group
delay are used, and it is the maximum, 18.4~mm, when the ionosphere path
delay reduction based on GPS TEC model is applied to \Note{the} X-band observables.

\begin{figure}
   \includegraphics[width=0.48\textwidth]{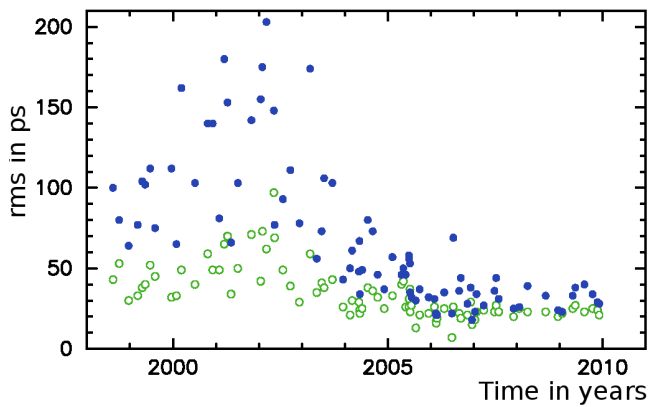}
   \caption{The rms of the ionosphere contribution from dual-band
            X/S observations (solid disks) and the rms of the differences
            of the path delay in the ionosphere from the dual-band
            X/S observations at the 1088~km long baseline
            \lavlba/\ovvlba\ and the GPS TEC ionosphere
            maps (circles).
           }
   \label{f:rms_iono_la_ov}
\end{figure}

  We compared these results with analogous observations in
the northern hemisphere. We picked the baseline \lavlba/\ovvlba\
that has almost exactly the same length as the baseline \hobart/\parkes:
1088~km against 1089~km. We processed 88 experiments at this baseline
in a similar fashion as we processed the \hobart/\parkes\ baseline.
Results are shown in Fig.~\ref{f:rms_iono_la_ov}. Applying
the ionosphere path delay from GPS has reduced the rms of post-fit
residuals at this baseline.

  Our interpretation of this phenomena is that the ionosphere path delay
from GPS TEC maps at baselines 1--2 thousands kilometres has the floor
around 30~ps. The dominant constituent in the ionospheric path delay
at these scales during the solar minimum are short-periodic scintillations
that the GPS TEC model with time resolution 2~hours and \Note{spatial
resolution} 500~km does not adequately represent. It is also known
\citep{r:gps-tec} that the accuracy of the GPS TEC model is worse
in the southern hemisphere than in the northern hemisphere due to the
disparity in the GPS station distribution.

  Applying the GPS TEC map ionosphere path delay reduction effectively
added noise in the data. Therefore, we did not apply the ionosphere path
delay in our final solution.

\subsubsection{Re-weighting observations}

  According to the Gauss-Markov theorem, the estimate of parameters
has the minimum dispersion when observation weights are chosen
reciprocal to the variance of errors. The group delays used in the
analysis have errors due to the thermal noise in fringe phases
and due to mismodeling the propagation delay:
\beq
      \sigma^2 = \sigma^2_{th} + \sigma^2_{na} + \sigma^2_{io} ,
\eeq{e:e2}
  where  $\sigma_{th}$ is the variance of the thermal noise,
$\sigma_{na}$, and $\sigma_{io}$ are the variances of errors of
modeling the path delay in the neutral atmosphere and the ionosphere
respectively.

  A rigorous analysis of the errors of modeling the path delay in
the neutral atmosphere is beyond the scope of this paper. Assuming
the dominant error source of the a~priori model is the high frequency
fluctuations of the water vapor at time scales less than 3--5 hours,
we sought a regression model for the dependence of $\sigma^2_{na}$ on the
non-hydrostatic component of the slanted path delay through the atmosphere.
We made several trial runs using all observing sessions under
geodesy and absolute astrometry programs for 30 years and added
in quadrature to the a~priori uncertainties of group delay an
additive correction:
\beq
   \sigma^2_{\rm used} = \sigma_{th}^2 + \lp a \cdot
                         \Frac{\tau_{s}}{\tau_z} \rp^2 ,
\eeq{e:e3}
  where ${\tau_s}$ is the contribution of the non-hydrostatic constituent
of the slanted path delay, ${\tau_z}$ is the non-hydrostatic path
delay in zenith direction computed by a direct integration of equations
of wave propagation through the atmosphere using the refractivity computed
using the MERRA model, and $a$ is a coefficient. We found that the baseline
length repeatability defined as the rms of the deviation of baseline length
with respect to the linear time evolution reaches the minimum when $a=0.02$.
Therefore, we adopted this value \Note{in} our analysis of LCS1 experiments.
For typical values of ${\tau_z}$, the added noise is 8~ps in zenith
direction and 16~ps at the elevation of $30^\circ$.

\begin{figure}
   \includegraphics[width=0.48\textwidth]{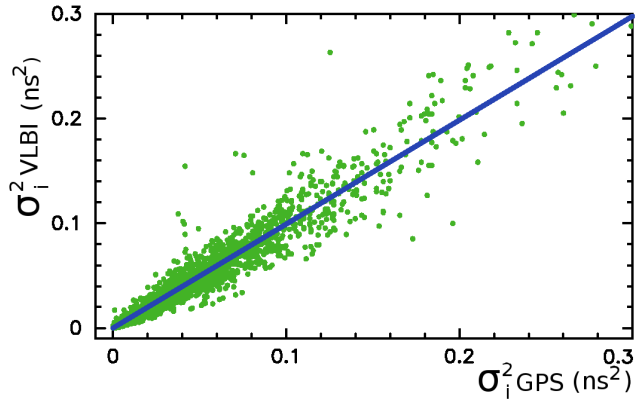}
   \caption{The dependence of the ionosphere path delay $\sigma^2_i$
            from dual-band VLBI versus $\sigma^2_i$ from TEC maps
            from GPS analysis. Data in time range 2005.0--2010.6
            (Solar minimum) were used.
           }
   \label{f:regr_iono_rms}
\end{figure}

   We computed the rms of ionosphere contributions from dual band VLBI
group delays and from GPS TEC maps. Fig.~\ref{f:regr_iono_rms} shows the
dependence of the square of the rms from dual-band VLBI versus
the square of the rms from GPS. The slope of the regression straight
line is $0.992 \pm 0.004$. This dependence suggests that although
the ionosphere path delay from TEC maps from GPS analysis is too
noisy to be applied for reduction of observations at short baselines during
the Solar minimum, it correctly predicts {\it the variance} of the
ionosphere path delay. We assigned the variance of the mismodeled
ionosphere path delay to the ionosphere contribution to delay computed
from TEC maps:
$\sigma_{io} = \tau_{io}$.

  We also computed, for each experiment and for each baseline, ad hoc
variances of observables that after being added in quadrature make the
ratio of the weighted sum of squares of post-fit residuals to their
mathematical expectation close to unity. This computation technique
is presented in \citet{r:vgaps}. The ad hoc variance was applied to
further inflate the formal uncertainties of {the observables that have}
already been corrected for the inaccuracy of the a~priori model of wave
propagation through the ionosphere and neutral atmosphere
(expression \ref{e:e2}). In contrast to $\sigma^2_{na}$ and
$\sigma^2_{io}$, the baseline-dependent ad~hoc variance is elevation
independent.

\subsubsection{Global LSQ solution}

\begin{figure}
   \includegraphics[width=0.48\textwidth]{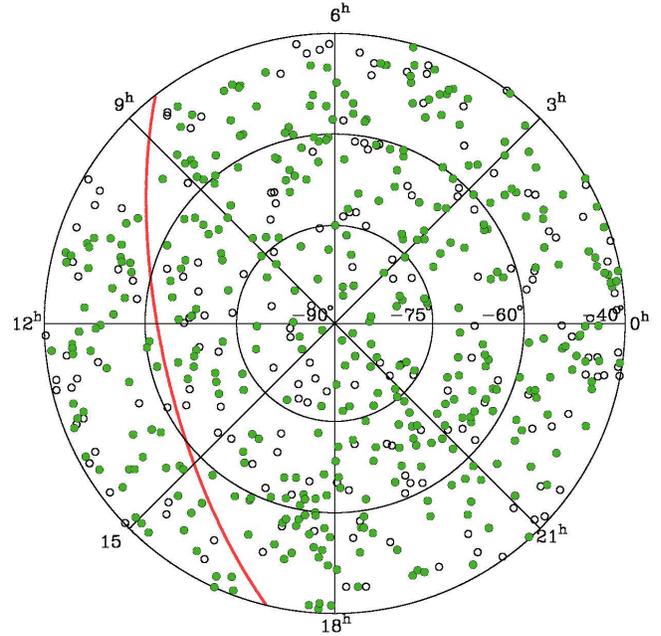}
   \caption{The distribution of sources with milliarcsecond
            positions at the southern sky. Circles denote
            sources known by 2008. Filled green discs
            denote sources from the LCS1 catalogue. The
            Galactic plane shown by red line.
           }
   \label{f:lcs1_map}
\end{figure}

  We ran a single global LSQ solution using all available dual-band VLBI
observations under geodesy and absolute astronomy programs from 1980.04.12
through 2010.08.04, in total 7.56 million observations, and the X-band VLBI
data from 4 LCS1 observing sessions. The RCP and LCP data were treated
as independent experiments. The following parameters were estimated over
the global dataset: coordinates of 4924 sources, including 410 detected
target objects in the LCS1 campaign (see Fig.~\ref{f:lcs1_map}); positions
and velocities of all stations; coefficients of expansion over the B-spline
basis of non-linear motions for 17 stations; coefficients of harmonic site
position variations of 48 stations at four frequencies: annual, semi-annual,
diurnal, semi-diurnal; and axis offsets for 67 stations. In addition, the
following parameters were estimated for each experiment
independently: station-dependent clock functions modeled by second order
polynomials, baseline-dependent clock offsets, the pole coordinates,
UT1-TAI, and daily nutation offset angles. The list of estimated parameters
also contained more than 1 million nuisance parameters: coefficients of
linear splines that model atmospheric path delay (20 minutes segment)
and clock function (60 minutes segment).

  The rate of change for the atmospheric path delay and clock function between
adjacent segments was constrained to zero with weights reciprocal to
$ 1.1 \cdot 10^{-14} $ and \mbox{$2\cdot10^{-14}$} respectively in order
to stabilize our solution. We apply no-net rotation constraints on position
of 212 sources marked as ``defining'' in the ICRF catalogue \citep{r:icrf98}
that requires the positions of these source in the new catalogue to have
no rotation with respect to the position in the ICRF catalogue.

  The global solution sets the orientation of the array with respect to
an ensemble of $\sim\! 5000$ extragalactic remote radio sources.
The orientation of that ensemble is defined by the series of the Earth
orientation parameters evaluated together with source coordinates.
Common 111 sources observed in LCS1 as atmosphere and amplitude
calibrators provide strong connection between the new catalogue to the
old catalogue of compact sources.

\subsection{Error analysis of the LCS1 catalogue}
\label{s:errors}

  To assess the systematic errors in our results we exploited the fact that
111 known sources were observed as amplitude and atmospheric calibrators.
Positions of these sources were determined from previous dual-band
S/X observations with accuracies better than 0.2~mas. We sorted the set
of 111 calibrators according to their right ascensions and split them into
two subsets of 55 and 56 objects, even and odd. We ran two additional
solutions. In the first solution we suppressed 55 calibrators in all
experiments but LCS1 and determined their positions solely from the LCS1
experiment. In the second solution we did the same with the second subset.
Considering that the positions of calibrators from numerous S/X
observations represent the ground truth, we treated the differences
as LCS1 errors.

  We computed the $\chi^2/\mbox{ndf}$ statistics for the differences in
right ascensions and declinations $\Delta_\alpha$ and $\Delta_\delta$ and
sought additional variances $v_\alpha$ and $v_\delta$ that, being added in
quadrature to the formal source position uncertainties, $\sigma_{\!\alpha,i}$
and $\sigma_{\!\delta,i}$, make them close to unity:
\beq
   \begin{array}{l @{\enskip} c @{\enskip} l}
      \frac{\chi^2_\alpha}{\mbox{ndf}} & = &
            \Frac{\sum \Delta \alpha_i^2 \cos^2\delta_i}
            {n \, \sum \sqrt{ \sigma^2_{\alpha,i} \cos^2 \delta_i +
                   v^2_\alpha} }
      \vex \\
      \frac{\chi^2_\delta}{\mbox{ndf}} & = & \Frac{\sum \Delta \delta_i^2 }
           {n \, \sum \sqrt{ \sigma^2_{\delta,i} + v^2_\delta} } .
   \end{array}
\eeq{e:e5}

  We found the following additive corrections of the uncertainties
in right ascensions scaled by $\cos \delta$ and for declinations
respectively: $v_\alpha = 1.44$~mas and $v_\delta$=0.51~mas. We do not
have an explanation why the additive correction for scaled right
ascensions is 3 times greater than for declinations. After applying
the additive corrections, the wrms of source position differences are
1.8~mas for right ascension scaled by $\cos\delta$ and 1.5~mas for
declinations.

  The final inflated \Note{uncertainties} of source positions,
$\sigma^2_\alpha(f)$ and $ \sigma^2_\delta(f)$, are
\beq
   \begin{array}{l @{\enskip} c @{\enskip} l}
      \sigma^2_\alpha(f) = \sigma^2_\alpha + v^2_\alpha/\cos^2 \delta
      \vex \\
      \sigma^2_\delta(f) = \sigma^2_\delta + v^2_\delta .
   \end{array}
\eeq{e:e6}

\subsection{Data analysis: correlated flux density determination}

  Each detected source has from 3 to 60 observations, with \Note{a median
value of 25}. Imaging a source with 25 points at the $uv$ plane is
a difficult problem, and the dynamic range of such images will be
between 1:10 and 1:100, which is far from spectacular. Images produced
with the hybrid self-calibration method will be presented in a separate
paper. In this study we limited our analysis \Note{to} mean correlated flux
density estimates in three ranges of lengths of the baseline projections
onto the plane tangential to the source, without inversion of calibrated
visibility data. Information about the correlated flux density is needed
for evaluation of the required integration time when an object is used
as a phase calibrator.

  First, we calibrated raw visibilities $v$ for the a~priori system
temperature $T_{sys}$ and antenna gain $G(e)$:
$ A_{corr} = v \cdot T_{sys}(t,e)/G(e)$. The coefficients of antenna gain
expansions into polynomials over elevation angle $e$ are presented
in Table~\ref{t:gain}. System temperature was measured at each station,
each scan.

\begin{table*}
  \caption{The coefficients of antenna gain polynomials over elevation angle
           expressed in radians. Gain = DPFU $\dss\sum_{k=0}^{k=5} a_k E^k$.
           The DFPU is shown in the second columns. Columns 3--8 hold
           coefficients of degree 0 through 5.}
  \label{t:gain}
  \begin{tabular}{llllllll}
      \hline
      Station        & DFPU K/Jy &
      \multicolumn{6}{l}{The coefficients of polynomial for gain as a function of the elevation
                         angle in degrees} \\
      && 0 \hphantom{aa} & 1 \hphantom{aaaa}  & \hphantom{aa}   2 &
           \hphantom{aa} 3 & \hphantom{aa} 4 & \hphantom{aa} 5 \\
      \hline
      \atca    &$ 0.100 $&$ 1.0     $&$ 0.0                 $&$ \hm 0.0                  $&$ \hm 0.0                 $&$ \hm 0.0                  $&$ \hm 0.0                  $ \\
      \ceduna  &$ 1.000 $&$ 0.9045  $&$ 4.032 \cdot 10^{-3} $&$    -4.280 \cdot 10^{-5}  $&$ \hm 0.0                 $&$ \hm 0.0                  $&$ \hm 0.0                  $ \\
      \tid     &$ 0.240 $&$ 0.9700  $&$ 2.927 \cdot 10^{-3} $&$    -1.070 \cdot 10^{-4}  $&$ \hm 1.761 \cdot 10^{-6} $&$    -1.112 \cdot 10^{-8 } $&$ \hm 0.0                  $ \\
      \hobart  &$ 0.058 $&$ 0.6997  $&$ 4.445 \cdot 10^{-2} $&$    -2.331 \cdot 10^{-3}  $&$ \hm 5.335 \cdot 10^{-5} $&$    -5.412 \cdot 10^{-7}  $&$ \hm 1.926 \cdot 10^{-9}  $ \\
      \mopra   &$ 0.095 $&$ 1.0     $&$ 0.0                 $&$ \hm 0.0                  $&$ \hm 0.0                 $&$ \hm 0.0                  $&$ \hm 0.0                  $ \\
      \parkes  &$ 0.480 $&$ 0.7900  $&$ 3.368 \cdot 10^{-3} $&$ \hm 8.507 \cdot 10^{-5 } $&$    -1.398 \cdot 10^{-6} $&$ \hm 0.0                  $&$ \hm 0.0                  $ \\
      \hline
  \end{tabular}
\end{table*}

  At the second step, we adjusted antenna gains using publicly available
brightness distributions of calibrator sources made with observations under
other programs.  We compiled The Database
of Brightness Distributions, Correlated Flux Densities and Images of Compact
Radio Sources Produced with
VLBI\footnote{Available at {\tt http://astrogeo.org/vlbi\_images}}
from authors who agreed to make their imaging results publicly available.
Among 111 sources observed as calibrators, images of 14--27 objects at
each experiment were available. These are images in the form of CLEAN
components mainly from \citet{r:sou_ima1,r:sou_ima2}.
CLEAN components of source brightness distributions from analysis
of the TANAMI program \citep{r:tanami} that observed with the LBA
concurrently with the LCS1 were not available, but the parameters of
one-component Gaussian models that fit the core regions were published.
For those sources for which both a set of CLEAN components
and the parameter of the Gaussian one-component model were available,
we used CLEAN components.

  We predicted the correlated flux density $F_{corr}$ for each observation
of an amplitude calibrator with known brightness distribution, as
\beq
   \begin{array}{lcl}
      F_{corr} & = & \left|
                   \dss\sum_i c_i e^{\frac{2\pi\, f}{c}\, (u\, x + v\, y)}
                   \right| \\
      F_{corr} & = & S \, e^{\frac{\pi^2}{4\ln2}\,
                      (a^2\, (u \cos\theta + v \sin\theta)^2 +
                       b^2\, (-u\sin\theta + v \cos\theta)^2)} ,
   \end{array}
\eeq{e:7}
   where $c_i$ is the correlated flux density of the $i$th CLEAN component
with coordinates $x$ and $y$ with respect to the center of the image;
$u$ and $v$ are the projections of the baseline vectors on the tangential
plane of the source; and $a$ and $b$ are the FWHM of the Gaussian that
approximates the core, and $\theta$ is the position angle of the semi-major
axis of the Gaussian model.

  Then we built a system of least square equations for all observations
of calibrators with known images used in astrometric solutions:
\beq
     F_{corr} = \sqrt{g_i \, g_j} A_{corr}
\eeq{e:8}
   and after taking logarithms from left and right hand sides solved
for corrections to gains $g$ for all stations. Finally, we applied
corrections to gain for observations of all other sources.

  The correlated flux density is a constant during observing session only
for unresolved sources. For resolved sources the correlated flux density
depends on the projection of the baseline vector on the source plane
and on its orientation. We binned the correlated flux densities in
three ranges of the baseline vector projection lengths, 0--6 M$\lambda$,
6--25 M$\lambda$, 25--50 M$\lambda$, and found the median value within
each bin. These bins correspond to scales of the detected emission at
$> 30$~mas, 7--30~mas, and $< 7$~mas respectively.

  \Note{The uncertainties} of our estimates of the correlated density depend on
the thermal noise described as  $\sigma_{th} = F_{corr}/\SNR$ and
errors of gain calibration. The variance of the thermal noise was in
the range of 1 to 6 mJy depending on the sensitivity of a baseline, with
the median value of 3~mJy. The LSQ solution for gains provided the variance
of the logarithm of gains. Assuming the calibration errors to be
multiplicative in the form $g \, (1 + \epsilon)$, where $g$ is gain
\Note{and $\epsilon$ is the Gaussian random variable, we can evaluate
the contribution of gain errors in the multiplicative uncertainty of
$F_{corr}(1 + \mu)$, where $\mu$ the Gaussian random variable.
Its variance is evaluated as
\beq
   \begin{array}{l}
     \sigma(\mu) = \\
                 \exp\lp \frac{1}{2}
                 \sqrt {\Cov(s_1,s_1) + \Cov(s_2,s_2) + 2\Cov(s_1,s_2) }
                 \rp ,
     \end{array}
\eeq{e:e9}
}
  where $\Cov(s_i,s_j)$ is the covariance of the logarithm of gain between
the $i$-th and $j$-th station. Multiplicative gain \Note{uncertainties}
are in the range 0.08--0.1 for \tid\ and 0.02--0.05 for other stations.
\Note{The gain uncertainties for
\tid\ are higher}, because it observed 3--5 times less than other stations.
The total variance is a sum in quadrature of $\sigma_c$ and $\sigma_{th}$.
We should note that our estimate of systematic errors does not account for
possible errors in gain curve determination. A systematic error in the gain
curve would directly affect our estimate of the correlated flux density.
It will also affect the maps of calibrators sources that we took from
literature, and thus indirectly affect our estimates of gain  corrections.
We do not have information about \Note{uncertainties} of gain curves
of LBA antennas.

\begin{table*}
   \caption{The first 12 rows of the catalogue of source correlated flux
            density estimates for both 410 target sources and 111 calibrator
            sources in each experiment. Some sources were observed
            on more than one experiment. The table columns are explained
            in the text. The full table is available in the electronic
            attachment.}
   \label{t:lcs1_fd}
   \begin{tabular}{llllllllllll}
     \hline
     J2000-name & B1950-name & Date & Sts & \# Obs
        & \nnntab{c}{Corr. flux density (Jy)}
        & \nnntab{c}{Flux density \note{uncertainties} (Jy)}
        & Exp \\
     \hline
     J0004$-$4736 & 0002$-$478 & 2008.11.28 &      &  12 &  0.352 & 0.268 & 0.267 &  0.015 & 0.019 & 0.013 & v271b \\
     J0011$-$8443 & 0009$-$850 & 2008.02.05 &      &  30 &  0.161 & 0.165 & 0.180 &  0.005 & 0.005 & 0.006 & v254b \\
     J0012$-$3954 & 0010$-$401 & 2008.02.05 &      &  16 &  0.761 & 0.769 & 0.715 &  0.016 & 0.012 & 0.014 & v254b \\
     J0012$-$3954 & 0010$-$401 & 2009.07.04 & Cal  &  10 &  0.815 & 0.654 & 0.640 &  0.026 & 0.023 & 0.028 & v271c \\
     J0028$-$7045 & 0026$-$710 & 2008.11.28 &      &  40 &  0.092 & 0.086 & 0.081 &  0.005 & 0.006 & 0.005 & v271b \\
     J0030$-$4224 & 0027$-$426 & 2009.07.04 &      &  30 &  0.312 & 0.290 & 0.256 &  0.011 & 0.010 & 0.011 & v271c \\
     J0033$-$4236 & 0031$-$428 & 2009.07.04 &      &  30 &  0.132 & 0.135 & 0.143 &  0.005 & 0.005 & 0.008 & v271c \\
     J0034$-$4116 & 0031$-$415 & 2009.07.04 &      &  18 &  0.053 & 0.045 & 0.039 &  0.003 & 0.002 & 0.004 & v271c \\
     J0040$-$4253 & 0038$-$431 & 2009.07.04 &      &  28 &  0.068 & 0.058 & 0.067 &  0.003 & 0.002 & 0.007 & v271c \\
     J0042$-$4030 & 0039$-$407 & 2009.07.04 &      &  26 &  0.255 & 0.253 & 0.276 &  0.010 & 0.011 & 0.013 & v271c \\
     J0042$-$4333 & 0040$-$438 & 2009.07.04 &      &  30 &  0.165 & 0.113 & 0.126 &  0.005 & 0.005 & 0.008 & v271c \\
     J0044$-$8422 & 0044$-$846 & 2008.02.05 &      &  26 &  0.278 & 0.255 & 0.234 &  0.007 & 0.006 & 0.007 & v254b \\
     \hline
   \end{tabular}
\end{table*}

  Table~\ref{t:lcs1_fd} displays 12 out of 633 rows of the catalogue of
correlated flux density estimates. The full table is available in the electronic
attachment. Column 1 and 2 contain the J2000 and B1950 IAU names of a source;
column 3 contains the observation date of the experiment; the 4th column
contains the status of the source: {\tt Cal} if it was used as amplitude
calibrator; the 5th columns contains the number of RCP observations used
in processing. Columns 6, 7, and 8 contain median values of correlated
flux densities determined in that experiment at baseline projection
lengths 0--6~$M\lambda$, 6--25~$M\lambda$, and 25--50~$M\lambda$ respectively.
Columns 9, 10, and 11 contain the arithmetic mean of correlated flux
density \Note{uncertainties} that accounts for both thermal noise and
uncertainties in calibration. If no data were collected to that
range of baseline projection lengths, $-1.0$ is used as a substitute.
Column 12 contains the experiment name.

\begin{figure}
   \includegraphics[width=0.48\textwidth]{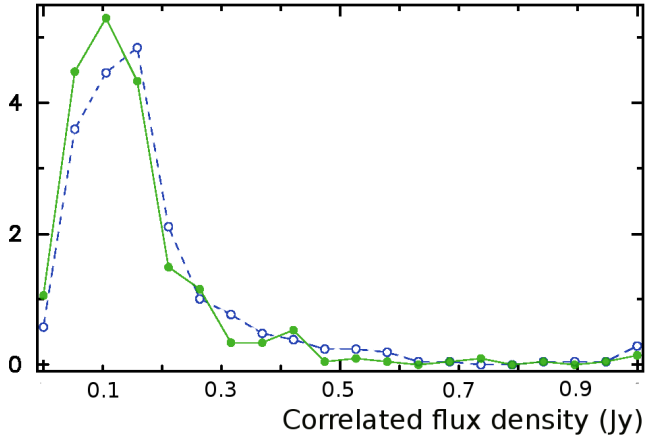}
   \caption{The probability density histogram of correlated flux
            densities in LCS1 experiments. The dotted line with
            dicks denotes correlated flux densities at baseline
            projection lengths 0--6 M$\lambda$, the solid line
            with discs denotes the correlated flux density at
            projection lengths 25--50 M$\lambda$,
           }
   \label{f:corr_hist}
\end{figure}

  Fig.~\ref{f:corr_hist} shows the probability density histogram of
correlated flux densities in LCS1 experiments.

\section{The LCS1 catalogue}
\label{s:catalog}

  Table~\ref{t:lcs1} displays 12 out of 410 rows of the LCS1 catalogue of
source positions. The full table is available in the electronic
attachment. Column 1 and 2 contain the J2000 and B1950 IAU names of a source;
column 3, 4, and 5 contain hours, minutes and seconds of the right ascension;
columns 6, 7, 8 contain degrees, minutes and arcseconds of declination.
Columns 9 and 10 contain inflated position \Note{uncertainties} in right
ascension
(without multiplier $\cos \delta$) and declination \Note{in milliarcseconds.
Column 11 lists the correlation coefficient between right ascension and
declination, \Note{and} column 12 contains the total number of observations used
in position data analysis, including RCP and LCP data. Columns 13, 14,
and 15 contain the median estimates of the correlated flux density in Jansky
over all experiment of the source in three ranges of baseline projection
lengths: 0--6~$M\lambda$, 6--25~$M\lambda$, and 25--50~$M\lambda$.
Columns 16, 17, and 18 contain estimates of the correlated flux density
uncertainties in Jansky.}

\begin{table*}
   \caption{The first 12 rows of the LCS1 catalogue of source positions
            of 410 target sources. The table columns are explained
            in the text. The full table is available in the electronic
            attachment.}
   \label{t:lcs1}
   \footnotesize
   \begin{tabular}{l @{\quad} l @{\:} r@{\:\:} l@{\:\:} l @{\!}
                   r @{\:} r @{\:}l
                   r@{\:\:\:}r@{\:\:\:}r @{\:\:\:} c @{\:\:\:}
                   l@{\:\:\:}l@{\:\:\:}l l@{\:\:\:}l@{\:\:\:}l}
      \hline
      J2000-name   & B1950-name & \nnntab{c}{Right ascension}
                   & \nnntab{c}{Declination}
                   & $\sigma(\alpha)$ & $ \sigma(\delta) $
                   & \ntab{c}{Corr} & \#Obs & \nnntab{c}{Correl. Flux}
                   & \nnntab{c}{Correl. Flux uncer.} \\
                   & & & & & & &
                   &   (9) & (10) & (11) & (12)
                   &  \nc{(13)} & \nc{(14)} & \nc{(15)} & \nc{(16)}
                   &  \nc{(17)} & \nc{(18)} \\
                   & & h & m & \hm s & ${}^\circ $ & $'$ & \hm $''$
                   &  mas  & mas  &      &
                   &  \nc{Jy} & \nc{Jy} & \nc{Jy} & \nc{Jy}
                   &  \nc{Jy} & \nc{Jy}\\
      \hline
      J0011$-$8443 & 0009$-$850 & 00 & 11 & 45.90267 & $-$84 & 43 & 20.0096 &  46.1 &   3.1 &  -0.228 &    30 &  0.162 & 0.165 & 0.180 &  0.005 & 0.005 & 0.006 \\
      J0028$-$7045 & 0026$-$710 & 00 & 28 & 41.56281 & $-$70 & 45 & 15.9267 &   7.1 &   1.7 &   0.006 &    40 &  0.092 & 0.086 & 0.081 &  0.005 & 0.006 & 0.005 \\
      J0030$-$4224 & 0027$-$426 & 00 & 30 & 17.49264 & $-$42 & 24 & 46.4827 &   2.0 &   1.4 &  -0.036 &    37 &  0.312 & 0.290 & 0.256 &  0.011 & 0.010 & 0.011 \\
      J0033$-$4236 & 0031$-$428 & 00 & 33 & 47.94145 & $-$42 & 36 & 14.0676 &   3.7 &   2.2 &  -0.012 &    37 &  0.132 & 0.135 & 0.143 &  0.005 & 0.005 & 0.008 \\
      J0034$-$4116 & 0031$-$415 & 00 & 34 & 04.40893 & $-$41 & 16 & 19.4729 &   7.1 &   3.6 &   0.089 &    25 &  0.053 & 0.045 & 0.039 &  0.003 & 0.002 & 0.004 \\
      J0040$-$4253 & 0038$-$431 & 00 & 40 & 32.51473 & $-$42 & 53 & 11.3916 &   4.5 &   2.8 &   0.057 &    35 &  0.068 & 0.058 & 0.067 &  0.003 & 0.002 & 0.007 \\
      J0042$-$4030 & 0039$-$407 & 00 & 42 & 01.22481 & $-$40 & 30 & 39.7419 &   3.4 &   2.0 &   0.013 &    33 &  0.255 & 0.253 & 0.276 &  0.010 & 0.011 & 0.013 \\
      J0042$-$4333 & 0040$-$438 & 00 & 42 & 24.86725 & $-$43 & 33 & 39.8164 &   3.9 &   2.3 &   0.002 &    37 &  0.165 & 0.113 & 0.126 &  0.005 & 0.005 & 0.008 \\
      J0044$-$8422 & 0044$-$846 & 00 & 44 & 26.68719 & $-$84 & 22 & 39.9895 &  44.0 &   2.9 &  -0.237 &    30 &  0.278 & 0.255 & 0.234 &  0.007 & 0.006 & 0.007 \\
      J0047$-$7530 & 0046$-$757 & 00 & 47 & 40.81228 & $-$75 & 30 & 11.3640 &   6.4 &   1.4 &  -0.631 &    41 &  0.139 & 0.118 & 0.078 &  0.006 & 0.008 & 0.006 \\
      J0049$-$4457 & 0046$-$452 & 00 & 49 & 16.62412 & $-$44 & 57 & 11.1658 &   3.7 &   2.1 &  -0.005 &    37 &  0.244 & 0.215 & 0.207 &  0.009 & 0.006 & 0.011 \\
      J0054$-$7534 & 0052$-$758 & 00 & 54 & 05.81337 & $-$75 & 34 & 03.6325 &   6.3 &   1.4 &  -0.803 &    44 &  0.094 & 0.087 & 0.080 &  0.004 & 0.007 & 0.006 \\
      \hline
   \end{tabular}
\end{table*}

\begin{figure}
   \includegraphics[width=0.48\textwidth]{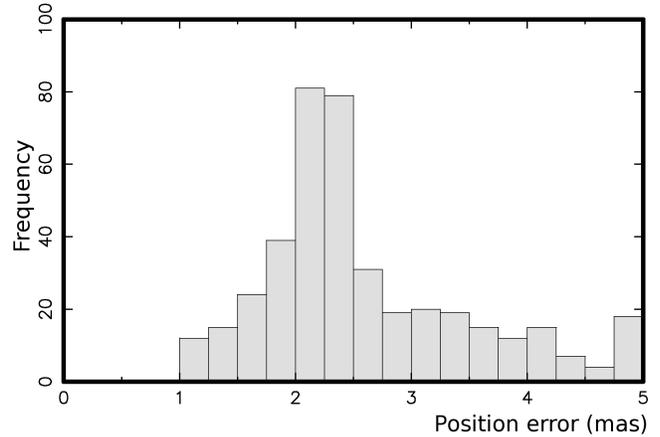}
   \caption{The histogram of the semi-major axes of inflated position
            error ellipses among 410 sources in the LCS1 catalogue.
           }
   \label{f:pos_err}
\end{figure}

  Of 421 sources observed, 410 were detected in three or more scans.
In four LCS1 observing sessions, 17,731 observations out of 19,494 were used
in the LSQ solution together with 7.56 million other VLBI observations.
The semi-major error ellipse of inflated position \Note{uncertainties}
varies in the range 1.4 to 16.8~mas with the median value of 2.6~mas.
The distribution of sources on the sky is presented in
Fig.~\ref{f:lcs1_map}. The histogram of position errors is shown in
Fig.~\ref{f:pos_err}.

\section{Summary}
\label{s:summary}

  The absolute astrometry LBA observations turned out highly successful.
The overall detection rate was 97\% --- the highest rate ever achieved
in a VLBI survey. If we exclude extended sources, non-AT20G sources and the
\Note{six} planetary nebulae \Note{included} in the candidate list due
\Note{to an} oversight, the detection rate is 99.8\%!
We attribute this high detection rate to two
factors. Firstly, the AT20G catalogue is highly reliable and is biased
\Note{towards very} compact objects. Selecting candidates based on
\Note{the}
simultaneous AT20G spectral index proves to be a good methodology. Secondly,
the LBA has very high sensitivity. The baseline detection limit over
2~minute of integration time varied from 7~mJy  to 30~mJy, with 7--12~mJy
at baselines with \parkes.

  We have successfully resolved group delay ambiguities with spacing 3.9~ns
for all observations. This became possible using the innovative algorithm
exploiting relatively low level of instrumental group delay errors.

  We have determined positions of 410 target sources never before observed
using VLBI, with \Note{a} median uncertainty \Note{of} 2.6~mas. Error analysis
showed a moderate contribution of the mismodeled ionosphere path delay to the
overall error budget. Both random and systematic errors are accounted  for
in the \Note{uncertainties} ascribed to source positions exploiting the
overlap of 111 additional sources observed in LCS1 experiments with their
positions known from prior observations. The positional accuracy of the LCS1
catalogue is a factor of 350 greater than the positional accuracy of the AT20G
catalogue, \Note{corresponding} to the ratio of the maximum baseline \Note{lengths}
of the LBA and the ATCA. The new catalogue has increased the number of sources
with declinations $< -40\degr$ from 158 to 568, i.e. by a factor of 3.5.

  We determined correlated flux densities for 410 target and 111
calibrator sources, and presented their median values in three ranges of
baseline projection lengths. The correlated flux density of \Note{the}
target sources varied from 0.02 to 2.5~Jy, \Note{and was in the range
80--300~mJy for 70\% of the} sources. \Note{The uncertainties} of the correlated
flux densities \Note{are estimated to be typically 5--8\%}.

  This observing program is continuing. By November 2010, four additional
twenty four hour experiments \Note{had been} observed with several more observing
sessions planned.

\section{Acknowledgments}
\label{s:acknowledgments}

\Note{The authors would like to thank Anastasios Tzioumis for comments
which helped to improve the manusrcipt.} The authors made use of the
database CATS of the Special Astrophysical Observatory. We used in our
work the dataset MAI6NPANA provided by the NASA/Global Modeling and
Assimilation Office (GMAO) in the framework of the MERRA atmospheric
reanalysis project. The Long Baseline Array is part of the Australia
Telescope National Facility which is funded by the Commonwealth of
Australia for operation as a National Facility managed by CSIRO.

\label{lastpage}

\end{document}